\documentstyle[12pt,linedraw]{article}
\newcommand{\tr}  {\rm{tr}}
\newcommand{\lsb}{{\scriptscriptstyle (}}
\newcommand{\rsb}{{\scriptscriptstyle )}}
\newcommand{\ssz} {\scriptstyle }
\newcommand{\Trns}{\scriptscriptstyle \rm T}
\newcommand{\be}  {\begin{equation}}
\newcommand{\ee}  {\end{equation}}

\begin{document}
\title{\bf Gluon -- W-meson scattering via different renormalization schemes.}

\author{M.M.Deminov${}^1$ and A.A.Slavnov${}^{1,2}$.}

\maketitle

\centerline{\it ${}^1$Moscow State University, Moscow, 117234, Russia.}
\centerline{\it ${}^2$Steklov Mathematical Institute, Russian Academy of Sciences,}
\centerline{\it Vavilova 42, Moscow, 117966, Russia.}

\begin{abstract}
The one loop gluon-W-meson amplitude is calculated
by means of the gauge-invariant generalized Pauli-Villars regularization
and with the help of dimensional regularization.
It is shown that in the former case the amplitude
satisfies Generalized Ward Identities,
whereas in the latter case the amplitude
differs from the first one
by the constant.
\end{abstract}

\section*{Introduction}

An important part of the calculations of radiative corrections in the Standard
model is a gauge invariant renormalization procedure.
The dimensional regularization which is mainly used for practical
calculations does not preserve chiral gauge invariance as
there is no self-consistent definition of $\gamma_5$ matrix in arbitrary dimension.
Of course one can use for calculations noninvariant regularization as well
and restore gauge invariance of renormalized theory
by adding noninvariant counterterms.
However in this case to fix renormalization freedom one
has to use Generalized Ward Identities (GWI). In the
case of the Standard model it complicates calculations considerably.
A good illustration of this problem is a calculation of
the one loop diagram describing the scattering of two gauge particles.
This diagram is superficially divergent but gauge invariance
fixes the renormalization freedom completely and imposing on the
amplitude GWI one gets unambiguous finite result.
This program was realized long ago for scattering
of light by light in QED \cite{KN1,KN2}.

In principle analogous procedure may be applied to the QCD process,
for example the gluon-W-meson amplitudes.
However in this case one has to use GWI related to the different gauge groups
(vectorial $\rm{SU(3)}$ and chiral $\rm{SU(2)}$), and contrary to the QED case
the amplitude in question is not symmetric with respect to
the interchange of all arguments. It makes the procedure quite cumbersome.

One may avoid those complications by using
a manifestly gauge invariant procedure.
In this case GWI are fulfilled automatically and one gets
an unambiguous result for this amplitude.

A gauge invariant regularization procedure for the Standard model
(generalized Pauli-Villars regularization or GPV) was proposed in \cite{main}.
In the present paper we apply this regularization to
the one loop amplitude describing the transition of two gluons to
two W-mesons.
We demonstrate that GPV regularization indeed produces
an unambiguous finite result which is consistent with GWI.
The formal using of the dimensional regularization
with $\gamma_5$-matrix defined as $\gamma_5 = i\gamma_0\gamma_1\gamma_2\gamma_3$
gives the amplitude which differs from the former one by a constant.

The following calculations were carried out for the Standard model.
To simplify the calculations we embed the Standard model
gauge group $\rm{SU(3) \times SU(2) \times U(1)}$ into the
the $\rm{SO(10)}$ group.

\section*{The generalized Pauli-Villars regularization}

The gauge invariant $\rm{SO(10)}$ Lagrangian, regularized with the help of
GPV~\cite{main}:
$$
{\cal L} =
-\frac{1}{4}F_{\mu\nu}^2 +
i \overline\psi_{+}\gamma_\mu (\partial_\mu - i g A^{ij}_\mu \sigma_{ij}) \psi_{+}
$$
\be
+ i \overline\varphi_r\gamma_\mu(\partial_\mu - i g A^{ij}_\mu \sigma_{ij}) \varphi_r
-\frac{1}{2} M_r\varphi_r^{\Trns} C_D C{\mit\Gamma}_{11}\varphi_r
-\frac{1}{2} M_r\overline\varphi_r C_D C{\mit\Gamma}_{11}\overline\varphi_r^{\Trns}
\label{SO10Main}
\ee
$$
+ i \overline\phi_r{\mit \Gamma}_{11}\gamma_\mu(\partial_\mu - i g A^{ij}_\mu \sigma_{ij}) \phi_r
-\frac{1}{2} M_r\phi_r^{\Trns} C_D C\phi_r
-\frac{1}{2} M_r\overline\phi_r C_D C\overline\phi_r^{\Trns}.
$$
Here the original spinors $\psi_{+}$ span the 16 dimensional representation of
$\rm{SO(10)}$ including quarks and leptons.
The spinor fields $\varphi_r$ are anticommuting and $\phi_r$ are commuting
Pauli-Villars fields with masses $M_r=Mr, r=1,\ldots,\infty.$ Each of $\varphi_r, \phi_r$
span the 32 dimensional representation of $\rm{SO(10).}$
The matrices $\sigma_{ij}$ are the $\rm{SO(10)}$ generators:
$\sigma_{ij} = \frac{i}{2}[\Gamma_i,\Gamma_j]$ where $\Gamma_i$ are
Hermitian $32$ by $32$ matrices which satisfy the Clifford algebra:
$\{\Gamma_i,\Gamma_j\} = 2\delta_{ij}$,
$\Gamma_{11} = {-i \Gamma_1 \Gamma_2 \ldots \Gamma_{10}}$.
The matrix $C$ is a conjugation matrix defined by relation
$\sigma^{\Trns}_{ij} C = - C \sigma_{ij}$.
The matrix $C_D$ is a charge conjugation matrix.
The Pauli-Villars mass $M$ should be taken infinite
after all calculations done.
All spinor fields have positive chirality with respect to the Lorentz group:
$\psi_{+} = \frac{1}{2}(1+\gamma_5)\psi_{+}$,
$\varphi = \frac{1}{2}(1+\gamma_5)\varphi$,
$\phi = \frac{1}{2}(1+\gamma_5)\phi$.
The 16 dimensional irreducible representations of SO(10) may be separated
with the help of the projection operators
$\frac{1}{2}(1 \pm \Gamma_{11})$:
$\psi_{+}=\frac{1}{2}(1+\Gamma_{11})\psi_{+}$,
$\varphi_\pm=\frac{1}{2}(1 \pm \Gamma_{11})\varphi$.
$\phi_\pm=\frac{1}{2}(1 \pm \Gamma_{11})\phi$.

Lagrangian~(\ref{SO10Main}) generates nonzero
propagators for \,
$\overline\varphi_\pm, \varphi_\pm;$\,
$\overline\varphi_\pm, \overline\varphi_\mp;$\,
$\varphi_\pm, \varphi_\mp;$
$\overline\phi_\pm, \phi_\pm;$\,
$\overline\phi_\pm, \overline\phi_\mp$ and
$\phi_\pm, \phi_\mp$.
They look as follows:
\be
S_{\overline\varphi_{r,+} \varphi_{r,+}}(k) =
S_{\overline\varphi_{r,-} \varphi_{r,-}}(k) =
S_{\overline\phi_{r,+} \phi_{r,+}}(k) =
S_{\overline\phi_{r,-} \phi_{r,-}}(k) =
\frac{\hat k}{k^2-M_r^2},
\label{Prop1}
\ee
\be
S_{\overline\varphi_{r,-} \overline\varphi_{r,+}}(k) =
S_{\varphi_{r,+} \varphi_{r,-}}(k) =
S_{\overline\phi_{r,-} \overline\phi_{r,+}}(k) =
S_{\phi_{r,+} \phi_{r,-}}(k) =
-\frac{M C_D C \Gamma_{11}}{k^2-M_r^2}.
\label{Prop2}
\ee

We construct the regularized four-point function
$G_{\mu \nu \lambda \sigma}(k^{\lsb {\ssz 1}\rsb},k^{\lsb {\ssz 2}\rsb},
                            k^{\lsb {\ssz 3}\rsb},k^{\lsb {\ssz 4}\rsb})$
corresponding to the amplitude of the process $W^+W^-\to gg.$
Here $k^{\lsb {\ssz 1,\ldots,4}\rsb}$ are momenta of the $W^+,W^-,g$ and
$g$ respectively.
In the following we often use a shorthand notation for
these momenta: $(k^{\lsb {\ssz 1}\rsb},k^{\lsb {\ssz 2}\rsb},k^{\lsb {\ssz 3}\rsb},k^{\lsb {\ssz 4}\rsb})$
$\to$ $(1,2,3,4).$

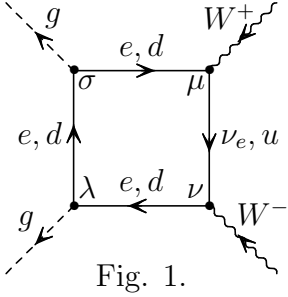
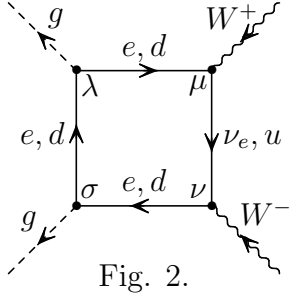
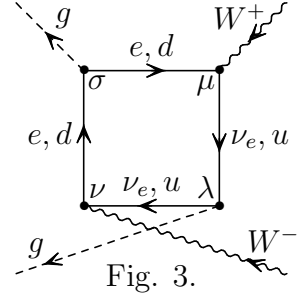
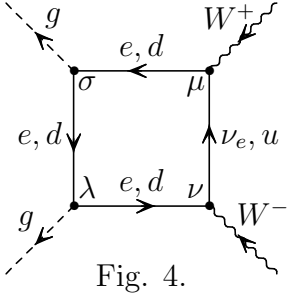
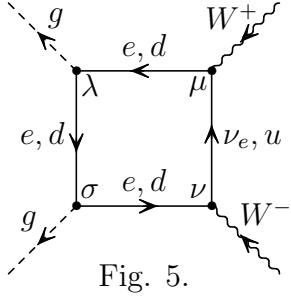
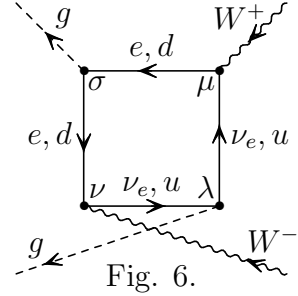
\begin{center}
\bigskip
\bigskip

\parbox{3.6cm}{%
\vbox{%
\hsize=4cm%
\Lengthunit=0.9cm%
\Linewidth{0.6pt}%
\GRAPH(hsize=5){\mov(1,0){\dashlin(-1,1)}%
\mov(0.5,0.5){\arrow0.05(-1.4,1.4)}%
\mov(1,0){\arrlin(2,0)}%
\mov(3,0){\arrlin(0,-2)}%
\mov(3,-2){\arrlin(-2,0)}%
\mov(1,-2){\arrlin(0,2)}%
\mov(3,0){\wavelin(1,1)}\mov(3.5,0.5){\arrow0.05(-1.4,-1.4)}%
\mov(3,-2){\wavelin(1,-1)}\mov(3.5,-2.5){\arrow0.05(-1.4,1.4)}%
\mov(1,-2){\dashlin(-1,-1)}\mov(0.5,-2.5){\arrow0.05(-1.4,-1.4)}%
\mov(1,0){\vdot\ind(2,-2){\sigma}}%
\mov(3,0){\vdot\ind(-2,-2){\mu}}%
\mov(3,-2){\vdot\ind(-2,2){\nu}}%
\mov(1,-2){\vdot\ind(2,2){\lambda}}%
\mov(0.6,0.6){\ind(1,2){g}}%
\mov(0.6,-2.6){\ind(-3,4){g}}%
\mov(3.4,-2.6){\ind(4,4){W^-}}%
\mov(3.4,0.6){\ind(-1,1){W^+}}%
\mov(2,0){\ind(0,3){e,d}}%
\mov(2,-2){\ind(0,3){e,d}}%
\mov(1,-1){\ind(-5,0){e,d}}%
\mov(3,-1){\ind(6,0){\nu_e,u}}%
\mov(2,-3.1){\ind(0,0){\mbox{Fig. 1.}}}}}}\quad%
\parbox{3.6cm}{%
\vbox{%
\hsize=3.6cm%
\Lengthunit=0.9cm%
\Linewidth{0.6pt}%
\GRAPH(hsize=5){\mov(1,0){\dashlin(-1,1)}%
\mov(0.5,0.5){\arrow0.05(-1.4,1.4)}%
\mov(1,0){\arrlin(2,0)}%
\mov(3,0){\arrlin(0,-2)}%
\mov(3,-2){\arrlin(-2,0)}%
\mov(1,-2){\arrlin(0,2)}%
\mov(3,0){\wavelin(1,1)}\mov(3.5,0.5){\arrow0.05(-1.4,-1.4)}%
\mov(3,-2){\wavelin(1,-1)}\mov(3.5,-2.5){\arrow0.05(-1.4,1.4)}%
\mov(1,-2){\dashlin(-1,-1)}\mov(0.5,-2.5){\arrow0.05(-1.4,-1.4)}%
\mov(1,0){\vdot\ind(2,-3){\lambda}}%
\mov(3,0){\vdot\ind(-2,-2){\mu}}%
\mov(3,-2){\vdot\ind(-2,2){\nu}}%
\mov(1,-2){\vdot\ind(2,2){\sigma}}%
\mov(0.6,0.6){\ind(1,2){g}}%
\mov(0.6,-2.6){\ind(-3,4){g}}%
\mov(3.4,-2.6){\ind(4,4){W^-}}%
\mov(3.4,0.6){\ind(-1,1){W^+}}%
\mov(2,0){\ind(0,3){e,d}}%
\mov(2,-2){\ind(0,3){e,d}}%
\mov(1,-1){\ind(-5,0){e,d}}%
\mov(3,-1){\ind(6,0){\nu_e,u}}%
\mov(2,-3.1){\ind(0,0){\mbox{Fig. 2.}}}}}}\quad%
\parbox{3.6cm}{%
\vbox{%
\hsize=3.6cm%
\Lengthunit=0.9cm%
\Linewidth{0.6pt}%
\GRAPH(hsize=5){\mov(1,0){\dashlin(-1,1)}%
\mov(0.5,0.5){\arrow0.05(-1.4,1.4)}%
\mov(1,0){\arrlin(2,0)}%
\mov(3,0){\arrlin(0,-2)}%
\mov(3,-2){\arrlin(-2,0)}%
\mov(1,-2){\arrlin(0,2)}%
\mov(3,0){\wavelin(1,1)}\mov(3.5,0.5){\arrow0.05(-1.4,-1.4)}%
\mov(1,-2){\wavelin(3,-1)}\mov(3.5,-2.834){\arrow0.05(-1.8,0.6)}%
\mov(3,-2){\dashlin(-3,-1)}\mov(0.5,-2.834){\arrow0.05(-1.8,-0.6)}%
\mov(1,0){\vdot\ind(2,-2){\sigma}}%
\mov(3,0){\vdot\ind(-2,-2){\mu}}%
\mov(3,-2){\vdot\ind(-2,2){\lambda}}%
\mov(1,-2){\vdot\ind(2,2){\nu}}%
\mov(0.6,0.6){\ind(1,2){g}}%
\mov(0.6,-2.6){\ind(-3,0){g}}%
\mov(3.4,-2.6){\ind(4,0){W^-}}%
\mov(3.4,0.6){\ind(-1,1){W^+}}%
\mov(2,0){\ind(0,3){e,d}}%
\mov(2,-2){\ind(0,3){\nu_e,u}}%
\mov(1,-1){\ind(-5,0){e,d}}%
\mov(3,-1){\ind(6,0){\nu_e,u}}%
\mov(2,-3.1){\ind(0,0){\mbox{Fig. 3.}}}}}}

\bigskip
\bigskip

\parbox{3.6cm}{%
\vbox{%
\hsize=3.6cm%
\Lengthunit=0.9cm%
\Linewidth{0.6pt}%
\GRAPH(hsize=5){\mov(1,0){\dashlin(-1,1)}%
\mov(0.5,0.5){\arrow0.05(-1.4,1.4)}%
\mov(1,0){\arrlin(0,-2)}%
\mov(3,0){\arrlin(-2,0)}%
\mov(3,-2){\arrlin(0,2)}%
\mov(1,-2){\arrlin(2,0)}%
\mov(3,0){\wavelin(1,1)}\mov(3.5,0.5){\arrow0.05(-1.4,-1.4)}%
\mov(3,-2){\wavelin(1,-1)}\mov(3.5,-2.5){\arrow0.05(-1.4,1.4)}%
\mov(1,-2){\dashlin(-1,-1)}\mov(0.5,-2.5){\arrow0.05(-1.4,-1.4)}%
\mov(1,0){\vdot\ind(2,-2){\sigma}}%
\mov(3,0){\vdot\ind(-2,-2){\mu}}%
\mov(3,-2){\vdot\ind(-2,2){\nu}}%
\mov(1,-2){\vdot\ind(2,2){\lambda}}%
\mov(0.6,0.6){\ind(1,2){g}}%
\mov(0.6,-2.6){\ind(-3,4){g}}%
\mov(3.4,-2.6){\ind(4,4){W^-}}%
\mov(3.4,0.6){\ind(-1,1){W^+}}%
\mov(2,0){\ind(0,3){e,d}}%
\mov(2,-2){\ind(0,3){e,d}}%
\mov(1,-1){\ind(-5,0){e,d}}%
\mov(3,-1){\ind(6,0){\nu_e,u}}%
\mov(2,-3.1){\ind(0,0){\mbox{Fig. 4.}}}}}}\quad%
\parbox{3.6cm}{%
\vbox{%
\hsize=3.6cm%
\Lengthunit=0.9cm%
\Linewidth{0.6pt}%
\GRAPH(hsize=5){\mov(1,0){\dashlin(-1,1)}%
\mov(0.5,0.5){\arrow0.05(-1.4,1.4)}%
\mov(1,0){\arrlin(0,-2)}%
\mov(3,0){\arrlin(-2,0)}%
\mov(3,-2){\arrlin(0,2)}%
\mov(1,-2){\arrlin(2,0)}%
\mov(3,0){\wavelin(1,1)}\mov(3.5,0.5){\arrow0.05(-1.4,-1.4)}%
\mov(3,-2){\wavelin(1,-1)}\mov(3.5,-2.5){\arrow0.05(-1.4,1.4)}%
\mov(1,-2){\dashlin(-1,-1)}\mov(0.5,-2.5){\arrow0.05(-1.4,-1.4)}%
\mov(1,0){\vdot\ind(2,-3){\lambda}}%
\mov(3,0){\vdot\ind(-2,-2){\mu}}%
\mov(3,-2){\vdot\ind(-2,2){\nu}}%
\mov(1,-2){\vdot\ind(2,2){\sigma}}%
\mov(0.6,0.6){\ind(1,2){g}}%
\mov(0.6,-2.6){\ind(-3,4){g}}%
\mov(3.4,-2.6){\ind(4,4){W^-}}%
\mov(3.4,0.6){\ind(-1,1){W^+}}%
\mov(2,0){\ind(0,3){e,d}}%
\mov(2,-2){\ind(0,3){e,d}}%
\mov(1,-1){\ind(-5,0){e,d}}%
\mov(3,-1){\ind(6,0){\nu_e,u}}%
\mov(2,-3.1){\ind(0,0){\mbox{Fig. 5.}}}}}}\quad%
\parbox{3.6cm}{%
\vbox{%
\hsize=3.6cm%
\Lengthunit=0.9cm%
\Linewidth{0.6pt}%
\GRAPH(hsize=5){\mov(1,0){\dashlin(-1,1)}%
\mov(0.5,0.5){\arrow0.05(-1.4,1.4)}%
\mov(1,0){\arrlin(0,-2)}%
\mov(3,0){\arrlin(-2,0)}%
\mov(3,-2){\arrlin(0,2)}%
\mov(1,-2){\arrlin(2,0)}%
\mov(3,0){\wavelin(1,1)}\mov(3.5,0.5){\arrow0.05(-1.4,-1.4)}%
\mov(1,-2){\wavelin(3,-1)}\mov(3.5,-2.834){\arrow0.05(-1.8,0.6)}%
\mov(3,-2){\dashlin(-3,-1)}\mov(0.5,-2.834){\arrow0.05(-1.8,-0.6)}%
\mov(1,0){\vdot\ind(2,-2){\sigma}}%
\mov(3,0){\vdot\ind(-2,-2){\mu}}%
\mov(3,-2){\vdot\ind(-2,2){\lambda}}%
\mov(1,-2){\vdot\ind(2,2){\nu}}%
\mov(0.6,0.6){\ind(1,2){g}}%
\mov(0.6,-2.6){\ind(-3,0){g}}%
\mov(3.4,-2.6){\ind(4,0){W^-}}%
\mov(3.4,0.6){\ind(-1,1){W^+}}%
\mov(2,0){\ind(0,3){e,d}}%
\mov(2,-2){\ind(0,3){\nu_e,u}}%
\mov(1,-1){\ind(-5,0){e,d}}%
\mov(3,-1){\ind(6,0){\nu_e,u}}%
\mov(2,-3.1){\ind(0,0){\mbox{Fig. 6.}}}}}}

\bigskip
\bigskip
\end{center}
Contribution of the original fields $\psi$ into the Feynman
diagram shown at Fig.1 is:
$$
I^{\lsb {\ssz 0}\rsb}{}_{\mu \nu \lambda \sigma}(1,2,3,4)=
\int{{\rm d}^{4}p \frac
{{\tr}\!\left[{\rm P_{+}}
\sigma^{\lsb {\ssz 1}\rsb}\gamma_\mu (\hat p+\hat k^{\lsb {\ssz 1}\rsb})\right.}
{((p+k^{\lsb {\ssz 1}\rsb})^2)}}{\scriptsize \times}
$$
\be
{\scriptsize \times}
\frac
{\left.
\sigma^{\lsb {\ssz 2}\rsb}\gamma_\nu (\hat p+\hat k^{\lsb {\ssz 1}\rsb} + \hat k^{\lsb {\ssz 2}\rsb})
\sigma^{\lsb {\ssz 3}\rsb}\gamma_\lambda (\hat p+\hat k^{\lsb {\ssz 1}\rsb}+\hat k^{\lsb {\ssz 2}\rsb}+\hat k^{\lsb {\ssz 3}\rsb})
\sigma^{\lsb {\ssz 4}\rsb}\gamma_\sigma \hat p )\right]}
{((p+k^{\lsb {\ssz 1}\rsb}+k^{\lsb {\ssz 2}\rsb})^2)((p+k^{\lsb {\ssz 1}\rsb}+k^{\lsb {\ssz 2}\rsb}+k^{\lsb {\ssz 3}\rsb})^2)(p^2)}.
\label{I0}
\ee
In this formula $\rm{P_+} =
\frac{1}{2}(1+\gamma_5)\frac{1}{2}(1+\Gamma_{11});$ \,
$\sigma^{\lsb {\ssz 1,\ldots,4}\rsb}$ are the linear combinations of
the $\rm{SO(10)}$-matrices $\sigma$ which correspond to the scattering particles.
Since ${\tr}(\Gamma_{11}$ $\sigma_{ij}$ $\sigma_{kl}$ $\sigma_{mn}$ $\sigma_{pr})$ $=$ $0$
the projector $\frac{1}{2}(1+\Gamma_{11})$ in (\ref{I0}) can be replaced by $1/2$.
Then the trace of the product of $\rm{SO}(10)$-matrices $32\times32$ from~(\ref{I0}) has the form:
\be
\frac{1}{2}{\tr}(\sigma^{\lsb {\ssz 1}\rsb}\sigma^{\lsb {\ssz 2}\rsb}\sigma^{\lsb {\ssz 3}\rsb}\sigma^{\lsb {\ssz 4}\rsb}).
\label{TrSigmas}
\ee
To study the process $W^+W^- \to gg$ we need the explicit form of $\rm{SU(2)}$ and $\rm{SU(3)}$ generators.
The $\rm{SU(3)}$ generator can be chosen
in the form $\sigma^{\lsb {\ssz 1}\rsb}=\sigma^{\lsb {\ssz 2}\rsb}=\frac{1}{2}(\sigma_{67}-\sigma_{58})$
and the $\rm{SU(2)}$ generators can be identified with
$\tau_k=\frac{1}{2}(\epsilon_{ijk}\sigma_{ij}-\sigma_{k4})$ (see~\cite{main}).
In the standard model mesons $W^\pm$ are the combinations of $\rm{SU(2)}$-gauge fields $W^{1,2,3}$:
$W^\pm=(W^1\mp iW^2)/\sqrt{2},$ and thus $\sigma^{\lsb {\ssz 1,2}\rsb} = (\tau_1\pm\tau_2)/\sqrt{2}.$
In the loop we are taking into account the contribution of the following virtual spinors only: $e^-,\nu_e,d,u.$
With the help of the $\sigma$'s algebra :
\be
\left[\sigma_{ij},\sigma_{kl}\right] =
i(\delta_{il}\sigma_{jk}-\delta_{ik}\sigma_{jl}
 +\delta_{jk}\sigma_{il}-\delta_{jl}\sigma_{ik}).
\label{SigmaAlgebra}
\ee
and keeping in mind the last remark it was calculated that the expression in~(\ref{TrSigmas})
equals to 1.

The contribution of the Pauli-Villars fields $\varphi_r,\phi_r$
is more complicated
due to the presence of the propagators:
$S_{\overline\varphi_{r,-} \overline\varphi_{r,+}}(k)$,
$S_{\varphi_{r,+} \varphi_{r,-}}(k)$,
$S_{\overline\phi_{r,-} \overline\phi_{r,+}}(k)$,
$S_{\phi_{r,+} \phi_{r,-}}(k)$.

If the propagators~(\ref{Prop1}) are denoted as $S$, and propagators~(\ref{Prop2})
-- as $\bar S$ then contributions of the Pauli-Villars fields can
be of four types:
$SSSS$, $S\bar SS\bar S$, $SS\bar S\bar S$, $\bar S\bar S\bar S\bar S$.
These contributions are denoted respectively as:
$I^{{\lsb} r{\ssz ,1,\pm}\rsb}$,
$I^{{\lsb} r{\ssz ,2,\pm}\rsb}$,
$I^{{\lsb} r{\ssz ,3,\pm}\rsb}$,
$I^{{\lsb} r{\ssz ,4,\pm}\rsb}$.
Sum of all these contributions is denoted as $I^{{\lsb} r,\pm{\rsb}}$ ($r \ge 1$):
\noindent
$I^{{\lsb} r{\ssz ,\pm}\rsb}_{\mu \nu \lambda \sigma}(1,2,3,4)=$
$$
=
 (I^{{\lsb} r{\ssz ,1,\pm}\rsb}_{\mu \nu \lambda \sigma}(1,2,3,4)
 +I^{{\lsb} r{\ssz ,2,\pm}\rsb}_{\mu \nu \lambda \sigma}(1,\ldots,4)
 +I^{{\lsb} r{\ssz ,3,\pm}\rsb}_{\mu \nu \lambda \sigma}(1,\ldots,4)
 +I^{{\lsb} r{\ssz ,4,\pm}\rsb}_{\mu \nu \lambda \sigma}(1,\ldots,4))=
$$
\be
=\int{
\frac
{{\rm d}^4p
 (\tilde I^{{\lsb} r{\ssz ,1,\pm}\rsb}{}_{\mu \nu \lambda \sigma}+
 \tilde I^{{\lsb} r{\ssz ,2,\pm}\rsb}{}_{\mu \nu \lambda \sigma}+
 \tilde I^{{\lsb} r{\ssz ,3,\pm}\rsb}{}_{\mu \nu \lambda \sigma}+
 \tilde I^{{\lsb} r{\ssz ,4,\pm}\rsb}{}_{\mu \nu \lambda \sigma})}
{((p+k^{\lsb {\ssz 1}\rsb})^2-M_r^2)((p+k^{\lsb {\ssz 1}\rsb}+k^{\lsb {\ssz 2}\rsb})^2-M_r^2)((p-k^{\lsb {\ssz 4}\rsb})^2-M_r^2)(p^2-M_r^2)}},
\label{IrInitial}
\ee
\noindent
where $\tilde I^{{\lsb} r,i,\pm\rsb}$ are the numerators of integrands of $I^{{\lsb} r,i,\pm\rsb}$:
\be
\tilde I^{{\lsb} r{\ssz ,1,\pm}\rsb}_{\mu \nu \lambda \sigma} =
{\tr}\!\left[{\rm P_\pm}
\sigma^{{\lsb}1{\rsb}}\gamma_\mu (\hat p\!+\!\hat k^{\lsb {\ssz 1}\rsb}) \sigma^{{\lsb}2{\rsb}}\gamma_\nu (\hat p\!+\!\hat k^{\lsb {\ssz 1}\rsb}\!+\!\hat k^{\lsb {\ssz 2}\rsb})
\sigma^{{\lsb}3{\rsb}}\gamma_\lambda (\hat p\!+\!\hat k^{\lsb {\ssz 1}\rsb}\!+\!\hat k^{\lsb {\ssz 2}\rsb}\!+\!\hat k^{\lsb {\ssz 3}\rsb}) \sigma^{{\lsb}4{\rsb}}\gamma_\sigma \hat p\right],
\label{Itildar1}
\ee
$$
\tilde I^{{\lsb} r{\ssz ,2,\pm}\rsb}_{\mu \nu \lambda \sigma} =
{\tr}\!\left[{\rm P_\pm} M_r^2
 \sigma^{{\lsb}1{\rsb}}\gamma_\mu (\hat p\!+\!\hat k^{\lsb {\ssz 1}\rsb})
 \sigma^{{\lsb}2{\rsb}}\gamma_\nu C_D C \Gamma_{11}
{\sigma^{{\lsb}3{\rsb}}}^{\Trns} \gamma_\lambda^{\Trns} (\hat p\!-\!\hat k^{\lsb {\ssz 4}\rsb})^{\Trns}
{\sigma^{{\lsb}4{\rsb}}}^{\Trns} \gamma_\sigma^{\Trns} C_D C \Gamma_{11}\right],
$$
\be
+{\tr}\!\left[{\rm P_\pm} M_r^2
 \sigma^{{\lsb}1{\rsb}}\gamma_\mu C_D C \Gamma_{11}
{\sigma^{{\lsb}2{\rsb}}}^{\Trns} \gamma_\nu^{\Trns} (\hat p\!+\!\hat k^{\lsb {\ssz 1}\rsb}\!+\!\hat k^{\lsb {\ssz 2}\rsb})^{\Trns}
{\sigma^{{\lsb}3{\rsb}}}^{\Trns} \gamma_\lambda^{\Trns} C_D C \Gamma_{11}
 \sigma^{{\lsb}4{\rsb}}\gamma_\sigma \hat p\right],
\label{Itildar2}
\ee
$$
\tilde I^{\lsb r{\ssz ,3,\pm}\rsb}_{\mu \nu \lambda \sigma} =
{\tr}\!\left[{\rm P_\pm} M_r^2
 \sigma^{{\lsb}1{\rsb}}\gamma_\mu (\hat p\!+\!\hat k^{\lsb {\ssz 1}\rsb})
 \sigma^{{\lsb}2{\rsb}}\gamma_\nu (\hat p\!+\!\hat k^{\lsb {\ssz 1}\rsb}\!+\!\hat k^{\lsb {\ssz 2}\rsb})
 \sigma^{{\lsb}3{\rsb}}\gamma_\lambda C_D C \Gamma_{11}
{\sigma^{{\lsb}4{\rsb}}}^{\Trns} \gamma_\sigma^{\Trns}
C_D C \Gamma_{11}\right]
$$
$$
+{\tr}\!\left[{\rm P_\pm} M_r^2
{\sigma^{{\lsb}1{\rsb}}}^{\Trns} \gamma_\mu^{\Trns} C_D C \Gamma_{11}
 \sigma^{{\lsb}2{\rsb}}\gamma_\nu (\hat p\!+\!\hat k^{\lsb {\ssz 1}\rsb}\!+\!\hat k^{\lsb {\ssz 2}\rsb})
 \sigma^{{\lsb}3{\rsb}}\gamma_\lambda (\hat p\!-\!\hat k^{\lsb {\ssz 4}\rsb})
 \sigma^{{\lsb}4{\rsb}}\gamma_\sigma C_D C \Gamma_{11}\right]
$$
$$
+{\tr}\!\left[{\rm P_\pm} M_r^2
 \sigma^{{\lsb}1{\rsb}}\gamma_\mu C_D C \Gamma_{11}
{\sigma^{{\lsb}2{\rsb}}}^{\Trns} \gamma_\nu^{\Trns} C_D C \Gamma_{11}
 \sigma^{{\lsb}3{\rsb}}\gamma_\lambda (\hat p\!+\!\hat k^{\lsb {\ssz 1}\rsb}\!+\!\hat k^{\lsb {\ssz 2}\rsb}\!+\!\hat k^{\lsb {\ssz 3}\rsb})
 \sigma^{{\lsb}4{\rsb}}\gamma_\sigma \hat p\right]
$$
\be
+{\tr}\!\left[{\rm P_\pm} M_r^2
 \sigma^{{\lsb}1{\rsb}}\gamma_\mu (\hat p\!+\!\hat k^{\lsb {\ssz 1}\rsb})
 \sigma^{{\lsb}2{\rsb}}\gamma_\nu C_D C \Gamma_{11}
{\sigma^{{\lsb}3{\rsb}}}^{\Trns} \gamma_\lambda^{\Trns} C_D C \Gamma_{11}
 \sigma^{{\lsb}4{\rsb}}\gamma_\sigma \hat p\right],
\label{Itildar3}
\ee
\be
\tilde I^{\lsb r{\ssz ,4,\pm}\rsb}_{\mu \nu \lambda \sigma} =
{\tr}\!\left[{\rm P_\pm} M_r^4
 \sigma^{{\lsb}1{\rsb}} \gamma_\mu C_D C \Gamma_{11}
{\sigma^{{\lsb}2{\rsb}}}^{\Trns} \gamma_\nu^{\Trns} C_D C \Gamma_{11}
 \sigma^{{\lsb}3{\rsb}}\gamma_\lambda C_D C \Gamma_{11}
{\sigma^{{\lsb}4{\rsb}}}^{\Trns} \gamma_\sigma^{\Trns}  C_D C \Gamma_{11}\right],
\label{Itildar4}
\ee
(here ${\rm P_\pm}=\frac{1}{2}(1\pm\Gamma_{11})\frac{1}{2}(1+\gamma_5)$).

\noindent
With the help of the identities:
$C^2=-1, C \Gamma_{11}=-\Gamma_{11} C, C_D \gamma_\mu = -\gamma_\mu^{\Trns} C_D,$
and
$
C_D C \Gamma_{11} {\sigma^{{\lsb}1{\rsb}}}^{\Trns} \gamma_\mu^{\Trns} \hat a^{\Trns} {\sigma^{{\lsb}2{\rsb}}}^{\Trns} \gamma_\nu^{\Trns} C_D C \Gamma_{11}
=-\sigma^{{\lsb}1{\rsb}} \gamma_\mu \hat a \sigma^{{\lsb}2{\rsb}} \gamma_\nu
$
the $I^{{\lsb} {\ssz r,\pm}\rsb}$ may be written in a more simple form (here ${\rm p_+}=\frac{1}{2}(1+\gamma_5)$):
$$
I^{\lsb r,\pm\rsb}_{\mu \nu \lambda \sigma}(k^{\lsb {\ssz 1}\rsb},k^{\lsb {\ssz 2}\rsb},k^{\lsb {\ssz 3}\rsb},k^{\lsb {\ssz 4}\rsb})=
{{\tr}(\frac{1\pm\Gamma_{11}}{2}\sigma^{{\lsb}1{\rsb}}\sigma^{{\lsb}2{\rsb}}\sigma^{{\lsb}3{\rsb}}\sigma^{{\lsb}4{\rsb}})}
\int{d^{4}p \frac
{{\tr}\!\left[{\rm p_+}
\gamma_\mu (\hat p + \hat k^{\lsb {\ssz 1}\rsb} - M_r)\right.}
{((p+k^{\lsb {\ssz 1}\rsb})^2-M_r^2)}} {\scriptstyle \times}
$$
\be
{\scriptstyle \times}
\frac
{\left.\gamma_\nu (\hat p + \hat k^{\lsb {\ssz 1}\rsb} + \hat k^{\lsb {\ssz 2}\rsb} - M_r)
\gamma_\lambda (\hat p + \hat k^{\lsb {\ssz 1}\rsb} + \hat k^{\lsb {\ssz 2}\rsb}+\hat k^{\lsb {\ssz 3}\rsb} -M_r) \gamma_\sigma (\hat p - M_r)\right]}
{((p+k^{\lsb {\ssz 1}\rsb}+k^{\lsb {\ssz 2}\rsb})^2-M_r^2)((p+k^{\lsb {\ssz 1}\rsb}+k^{\lsb {\ssz 2}\rsb}+k^{\lsb {\ssz 3}\rsb})^2-M_r^2)(p^2-M_r^2)}.
\label{Ir}
\ee
One may note that the term, proportional to $\Gamma_{11}$ vanishes, and
$I^{{\lsb} r,+\rsb} = I^{{\lsb} r,-\rsb}$.
The expression for the $I^{\lsb {\ssz 0}\rsb}$ in~(\ref{I0})
is the particular case of the $I^{\lsb r,+\rsb}$ (when $r=0$).

For explicit calculations Feynman parametrization is used:
$\alpha + \beta + \gamma + \delta = 1$,
and the integration over the Feynman parameters is defined as follows:
\be
\frac{1}{6}\int\limits_0^1{d\alpha \int\limits_0^{1-\alpha}{d\beta \int\limits_0^{1-\alpha-\beta}{d\gamma}}}
\equiv \int{d\tau}.
\ee
\noindent
For simplicity integration variable in the
$\Pi (k^{\lsb {\ssz 1}\rsb},k^{\lsb {\ssz 2}\rsb},k^{\lsb {\ssz 3}\rsb},k^{\lsb {\ssz 4}\rsb})$
is shifted:
\be
p \to p-(
\alpha  k^{\lsb {\ssz 1}\rsb} +
\beta  (k^{\lsb {\ssz 1}\rsb} + k^{\lsb {\ssz 2}\rsb})+
\gamma (k^{\lsb {\ssz 1}\rsb} + k^{\lsb {\ssz 2}\rsb} + k^{\lsb {\ssz 3}\rsb})),
\ee
\be
w \equiv
\alpha  k^{\lsb {\ssz 1}\rsb} +
\beta  (k^{\lsb {\ssz 1}\rsb} + k^{\lsb {\ssz 2}\rsb}) +
\gamma (k^{\lsb {\ssz 1}\rsb} + k^{\lsb {\ssz 2}\rsb} + k^{\lsb {\ssz 3}\rsb})
\ee
Thus $I^{{\lsb} r\rsb}_{\mu \nu \lambda \sigma}(1,2,3,4)$
can be rewritten in the form:
\be
\int{d\tau \int{\frac
{d^{4}p\quad{\tr}(\frac{1}{2}(1+\gamma_5)
\gamma_\mu (\hat p + k^{\lsb {\ssz 1}\rsb} - \hat w - M_r)
\ldots
\gamma_\sigma (\hat p - \hat w -M_r))}
{(p^2 - M_r^2 +
\alpha  (k^{\lsb {\ssz 1}\rsb})^2 +
\beta   (k^{\lsb {\ssz 1}\rsb} + k^{\lsb {\ssz 2}\rsb})^2 +
\gamma  (k^{\lsb {\ssz 1}\rsb} + k^{\lsb {\ssz 2}\rsb} + k^{\lsb {\ssz 3}\rsb}))^2
-w^2 )^4}}}.
\ee
\noindent
After this shift, the terms in the numerator proportional to
$p$ and $p^3$ may be omitted.
The denominator can be represented as:
$$
(p^2+
 (k^{\lsb {\ssz 1}\rsb})^2 \alpha \delta + (k^{\lsb {\ssz 2}\rsb})^2 \alpha \beta +
 (k^{\lsb {\ssz 3}\rsb})^2 \beta \gamma  + (k^{\lsb {\ssz 4}\rsb})^2 \gamma \delta -
$$
$$
 (k^{\lsb {\ssz 1}\rsb} + k^{\lsb {\ssz 2}\rsb}) (k^{\lsb {\ssz 3}\rsb} + k^{\lsb {\ssz 4}\rsb}) \beta \delta -
 (k^{\lsb {\ssz 1}\rsb} + k^{\lsb {\ssz 4}\rsb}) (k^{\lsb {\ssz 2}\rsb} + k^{\lsb {\ssz 3}\rsb}) \alpha \gamma - M_r^2)^4 \equiv
$$
\be
\equiv (p^2 + F(\alpha ,\beta ,\gamma ,\delta ,1,2,3,4,M_r^2))^4.
\ee
Regularized expression for the Feynman diagram shown at Fig.1 looks
as follows:
\be
\Pi_{\mu \nu \lambda \sigma}(1,2,3,4)=
I^{\lsb 0\rsb}_{\mu \nu \lambda \sigma}(1,2,3,4)+
\sum_{r=1}^{+\infty}{(-1)^r
\left(I^{{\lsb} r,+{\rsb}}_{\mu \nu \lambda \sigma}(1,2,3,4)+
      I^{{\lsb} r,-{\rsb}}_{\mu \nu \lambda \sigma}(1,2,3,4)\right)}.
\label{Pi}
\ee
The integrands can be represented as linear combinations of the following expressions:
\be
\sum_{r=-\infty}^{+\infty} \frac{r^{0,2,4}(-1)^r}{(p^2-M^2r^2+F)^4},
\label{Conv1}
\ee
which can be rewritten as derivatives of
$\sum\limits_{r=-\infty}^{+\infty} \frac{(-1)^r}{(p^2-M^2r^2+F)}$ over $p^2$ and $M^2$.
The following equation holds:
\be
\sum_{r=-\infty}^{+\infty} \frac{(-1)^r}{p^2+F-M^2r^2}=
\frac{\pi}{\sqrt{M^2(p^2+F)}\sin (\pi\sqrt{(p^2+F)/M^2})}.
\label{Sumr}
\ee
Since after the Wick rotation $p^2+F$ becomes negative,
one sees that integrand of $\Pi$ is proportional to:
\be
\frac{\pi}{M\sqrt{|p^2+F|}\rm{sinh} (\pi\sqrt{|p^2+F|}/M)}.
\label{PiSim}
\ee
This expression decreases rapidly providing the regularization.
But eq.~(\ref{PiSim}) is too complicated for further calculations.

Equation~(\ref{PiSim}) arises after summation of contributions
of an infinie series of PV fields.
The necessity to introduce an infinite system of PV fields is due
to the fact that these fields belong to the 32-dimensional
representation of $\rm{SO(10)}$ and they contribute to the
amplitude under consideration both $I^{{\lsb}r,+{\rsb}}$ and
$I^{{\lsb}r,-{\rsb}}$ whereas the original fields $\psi$
contribute only $I^{{\lsb}0,+{\rsb}}$. For this reason as was
discussed in ref.~\cite{main}, a finite number of PV fields does not
allow to satisfy the PV conditions providing the convergence of
regularized amplitude. To construct a Lagrangian gauge invariant
regularization for the Standard model with odd number of
generations an infinite set of PV fields is needed.
However being interested in the calculation of a particular
four-point amplitude we can omit the terms $I^{{\lsb}r,-{\rsb}}$
without spoiling the gauge invariance.
(As for this amplitude $I^{{\lsb}r,-{\rsb}}$ $=$ $I^{{\lsb}r,+{\rsb}}$
it corresponds to replacing the determinant which arise after
integration over the PV fields by the square root of the
determinant.) Then to regularize the model it is sufficient to
take a finite number of PV fields (r=1).

So instead of~(\ref{Pi}) the following formula for $\Pi$ may be used:
\be
\Pi_{\mu \nu \lambda \sigma}(1,2,3,4)=
I^{\lsb {\ssz 0}\rsb}{}_{\mu \nu \lambda \sigma}(1,2,3,4)-
I^{\lsb {\ssz 1,+}\rsb}{}_{\mu \nu \lambda \sigma}(1,2,3,4).
\label{PiTwo}
\ee
To illustrate this statement let us note that the integrand of the difference
between~(\ref{Pi}) and~(\ref{PiTwo}) is proportional to the following:
\be
\int{
\sum_{r=-\infty}^{+\infty} \frac{(-1)^r}{(p^2+F-M^2-M^2r^2)^4}},
\label{ToBeZero}
\ee
thus the difference between~(\ref{Pi}) and~(\ref{PiTwo}) tends to zero
when Pauli-Villars mass $M \to \infty$, and the gauge invariance is preserved.

The complete four-point function $G$ is the sum of $\Pi$'s
corresponding to the diagrams at Fig.~$1,\ldots,6$,
which differ by permutations of $W^+, W^-$ and $g$:
$$
G_{\mu \nu \lambda \sigma}(k^{\lsb {\ssz 1}\rsb},k^{\lsb {\ssz 2}\rsb},k^{\lsb {\ssz 3}\rsb},k^{\lsb {\ssz 4}\rsb})=
$$
\be
\Pi_{\mu \nu \lambda \sigma}(1,2,3,4)+
\Pi_{\mu \nu \sigma \lambda}(1,2,4,3)+
\Pi_{\mu \lambda \nu \sigma}(1,3,2,4)+
\label{GSix}
\ee
$$
+\Pi_{\mu \sigma \lambda \nu}(1,4,3,2)+
 \Pi_{\mu \lambda \sigma \nu}(1,3,4,2)+
 \Pi_{\mu \sigma \nu \lambda}(1,4,2,3).
$$

Below we demonstrate that this amplitude satisfies generalized
Ward Identities and in the limit $M\to\infty$ leads to a finite
gauge invariant expression.

The GWI for the four-point function
$G_{\mu \nu \lambda \sigma}(k^{\lsb {\ssz 1}\rsb},k^{\lsb {\ssz 2}\rsb},k^{\lsb {\ssz 3}\rsb},k^{\lsb {\ssz 4}\rsb})$
look as follows:
$$
k^{\lsb {\ssz 1}\rsb}_\mu G_{\mu \nu \lambda \sigma} (k^{\lsb {\ssz 1}\rsb},k^{\lsb {\ssz 2}\rsb},k^{\lsb {\ssz 3}\rsb},k^{\lsb {\ssz 4}\rsb})
+\left[ A_\mu(k^{\lsb {\ssz 1}\rsb}), G_{\mu \nu \lambda \sigma} (k^{\lsb {\ssz 1}\rsb},k^{\lsb {\ssz 2}\rsb},k^{\lsb {\ssz 3}\rsb},k^{\lsb {\ssz 4}\rsb})\right] =0,
$$
\be
k^{\lsb {\ssz 2}\rsb}_\nu G_{\mu \nu \lambda \sigma} (k^{\lsb {\ssz 1}\rsb},k^{\lsb {\ssz 2}\rsb},k^{\lsb {\ssz 3}\rsb},k^{\lsb {\ssz 4}\rsb})
+\left[ A_\nu(k^{\lsb {\ssz 2}\rsb}), G_{\mu \nu \lambda \sigma} (k^{\lsb {\ssz 1}\rsb},k^{\lsb {\ssz 2}\rsb},k^{\lsb {\ssz 3}\rsb},k^{\lsb {\ssz 4}\rsb})\right] =0, \quad \mbox{etc.}
\label{WardIdOff}
\ee
They are more complicated than in QED and in general their
analysis is quite cumbersome. However for the fourth order
amplitude we are considering, the terms proportional to $A_\mu$ do
not contribute and the GWI acquire a simple form:
\be
k^{\lsb {\ssz 1}\rsb}_\mu G_{\mu \nu \lambda \sigma}(1,2,3,4)=0, \qquad
\quad
k^{\lsb {\ssz 2}\rsb}_\nu G_{\mu \nu \lambda \sigma}(1,2,3,4)=0,
\quad\mbox{etc.}
\label{WardIdOn}
\ee
These identities differ from QED only by the absence of symmetry
with respect to the interchange of the momenta. Moreover, one can
show that the fourth order fermion loop diagram have to satisfy
the eq.~(\ref{WardIdOn}) also in the model with spontaneously
broken symmetry, when the fermions acquire nonzero masses. In
principle GWI in this case will include a contribution of the
Higgs fields, but for the diagram under consideration this
contribution vanishes.

For this reason in the following we consider a general case when
the fermions are massive. The physical fermion propagator looks
as follows:
\be
S_{\bar\psi\psi}=\frac{\hat k - m}{(k^2-m^2)},
\label{Propm}
\ee
and the interaction vertices include the chiral projector.

One can note that
$I(k^{\lsb {\ssz i}\rsb},k^{\lsb {\ssz j}\rsb},k^{\lsb {\ssz k}\rsb},k^{\lsb {\ssz l}\rsb})
=I(-k^{\lsb {\ssz i}\rsb},-k^{\lsb {\ssz j}\rsb},-k^{\lsb {\ssz k}\rsb},-k^{\lsb {\ssz l}\rsb}).$
Thus terms proportional to ${\tr}(\gamma_5 \ldots)$
in $G$ cancel in pairs identically. For example,
such contributions into $I_{\mu \sigma \lambda \nu}(1,4,3,2)$
and $I_{\mu \nu \lambda \sigma}(1,2,3,4)$ cancel each other:
$$
{\tr}(
\gamma_5
\gamma_\mu
(\hat p + \hat k^{\lsb {\ssz 1}\rsb} - m)
\gamma_\sigma
(\hat p + \hat k^{\lsb {\ssz 1}\rsb} + \hat k^{\lsb {\ssz 4}\rsb} - m)
\gamma_\lambda
(\hat p + \hat k^{\lsb {\ssz 1}\rsb} + \hat k^{\lsb {\ssz 4}\rsb} + \hat k^{\lsb {\ssz 3}\rsb} - m)
\gamma_\nu
(\hat p - m))=
$$
$$
=
-{\tr}(
\gamma_5
\gamma_\mu
(\hat p - \hat k^{\lsb {\ssz 1}\rsb} + m)
\gamma_\nu
(\hat p - \hat k^{\lsb {\ssz 1}\rsb} - \hat k^{\lsb {\ssz 2}\rsb}  + m)
\gamma_\lambda
(\hat p - \hat k^{\lsb {\ssz 1}\rsb} - \hat k^{\lsb {\ssz 2}\rsb} - \hat k^{\lsb {\ssz 3}\rsb}  + m)
\gamma_\sigma
(\hat p  + m))
$$
Thus the expression for this four-point function $G$ is equal to
the expression for the four-point function in non-chiral theory multiplied by $1/2$.
Therefore one expects that the fourth order scattering amplitude
calculated with the use of GPV regularization is gauge invariant.

An explicit check of the GWI for the GPV regularized four-point function can be carried out.
Let us note that since $I^{\lsb {\ssz 0}\rsb}_{\mu \nu \lambda \sigma}(1,2,3,4)-I^{\lsb {\ssz 1,+}\rsb}_{\mu \nu \lambda \sigma}(1,2,3,4)$
converges as $\int{{\rm d}^4p/((p^2)^3)}$ the integrand can be multiplied by
$(p_\mu - m)$ $-$ $(p_\mu + k^{\lsb {\ssz 2}\rsb}_\mu + k^{\lsb {\ssz 3}\rsb}_\mu + k^{\lsb {\ssz 4}\rsb}_\mu - m)$
$=k^{\lsb {\ssz 1}\rsb}_\mu$, and thus

\noindent
$k_\mu^{\lsb {\ssz 1}\rsb}
(I^{\lsb {\ssz 0}\rsb}_{\mu \nu \lambda \sigma}(1,2,3,4)-I^{\lsb {\ssz 1}\rsb}_{\mu \nu \lambda \sigma}(1,2,3,4))=$
$$
\left[
\int{d^{4}p
\frac{1}{2}
\left(
\frac
{{\tr}(\gamma_\nu (\hat p + \hat k^{\lsb {\ssz 3}\rsb} + \hat k^{\lsb {\ssz 4}\rsb} - m)
\gamma_\lambda (\hat p + \hat k^{\lsb {\ssz 4}\rsb} -m) \gamma_\sigma (\hat p - m))}
{((p + k^{\lsb {\ssz 3}\rsb} + k^{\lsb {\ssz 4}\rsb})^2-m^2)((p + k^{\lsb {\ssz 4}\rsb})^2-m^2)(p^2-m^2)}
\right.}
\right.
$$
$$
\left.
\left.-\frac
{{\tr}(\gamma_\nu (\hat p + \hat k^{\lsb {\ssz 3}\rsb} + \hat k^{\lsb {\ssz 4}\rsb} - m)
\gamma_\lambda (\hat p + \hat k^{\lsb {\ssz 4}\rsb} -m) \gamma_\sigma (\hat p + \hat k^{\lsb {\ssz 1}\rsb} - m))}
{((p-k^{\lsb {\ssz 3}\rsb}-k^{\lsb {\ssz 4}\rsb})^2-m^2)((p-k^{\lsb {\ssz 4}\rsb})^2-m^2)((p - \hat k^{\lsb {\ssz 1}\rsb})^2-m^2)}\right)
\right]
-\left.{\left[{\LARGE{\;}}\right]}\right|_{m \to M} \equiv
$$
$$
\equiv [A^{{\lsb}m{\rsb}}_{\nu \lambda \sigma}(1,2,3,4)-B^{{\lsb}m{\rsb}}_{\nu\lambda\sigma}(1,2,3,4)]-
[A^{{\lsb}M{\rsb}}_{\nu \lambda\sigma}(1,2,3,4)-B^{{\lsb}M{\rsb}}_{\nu\lambda\sigma}(1,2,3,4)].
$$
The sum of these $A$'s and $B$'s is zero.
For example, $A^{{\lsb}m{\rsb}}_{\nu \lambda \sigma}(1,2,3,4)$ cancel
$B^{{\lsb}m{\rsb}}_{\lambda \sigma \nu}(1,$ $4,$ $2,$ $3)$ ($p$ in the $B^{{\lsb}m{\rsb}}_{\lambda \sigma \nu}$ should be shifted:
$p \to p + k^{\lsb {\ssz 2}\rsb}$).
Other GWI~(\ref{WardIdOn}) are checked in the same way.

As it was expected the scattering amplitude calculated with the help of GPV method
satisfy automatically generalized Ward Identities.
In what follows this regularization method
will be compared with other methods.

\section*{The renormalization with the help of the Generalized Ward Identities (GWI)}

The only term in the numerator of the integrand of $I^{{\lsb}0{\rsb}}$,
which can lead to divergency is:
\be
{\tr}(\frac{1}{2}
\gamma_\mu \hat p
\gamma_\nu \hat p
\gamma_\lambda \hat p
\gamma_\sigma \hat p).
\label{NumDiv}
\ee
(In what follows we use the notation $I^{{\lsb}0{\rsb}}$ $\equiv$ $I$.)
The part of $I$, which contains only this term in the numerator
will be called as the 'potentially divergent part'.

One may note that for some  $\mu, \nu, \lambda, \sigma$ the
potentially divergent part of $I$ vanishes.
With the help of the identity:
\be
\int{p_\alpha p_\beta p_\gamma p_\delta f(p^2){\rm d}^dp } =
\frac{1}{d(d+2)}
(\delta_{\alpha\beta} \delta_{\gamma\delta}
+\delta_{\alpha\gamma}\delta_{\beta\delta}
+\delta_{\alpha\delta}\delta_{\beta\gamma})
\int{(p^2)^2 f(p^2){\rm d}^dp }.
\label{IntpId}
\ee
this potentially divergent part of $I_{\mu\nu\lambda\sigma}$ can be represented as follows:
$$
\int{
{\tr}(
\gamma_\mu \hat p
\gamma_\nu \hat p
\gamma_\lambda \hat p
\gamma_\sigma \hat p)
f(p^2) {\rm d}^4p}.
$$
Taking trace over spinoral indices one gets the following result:
\be
\frac{4}{3}(\delta_{\mu\nu}\delta_{\lambda\sigma}+\delta_{\mu\sigma}\delta_{\nu\lambda}-2\delta_{\mu\lambda}\delta_{\nu\sigma})
\int{(p^2)^2f(p^2) {\rm d}^4p}.
\label{IntpDiv2}
\ee
If
$ \delta_{\mu\nu}\delta_{\lambda\sigma}
 +\delta_{\mu\sigma}\delta_{\nu\lambda}
-2\delta_{\mu\lambda}\delta_{\nu\sigma}=0$,
then $I_{\mu\nu\lambda\sigma}$ is finite and does not need renormalization.

The set of indices $\{\mu, \nu, \lambda, \sigma\}$ for which the
expression in~(\ref{IntpDiv2}) is zero (i.e. $I_{\mu\nu\lambda\sigma}$ -- finite)
is large enough for finding $I$ with arbitrary indices with the help of the
GWI~(\ref{WardIdOn}).

Indeed, let the indices of a potentially divergent diagram satisfy the condition:
$ \delta_{\mu\nu}\delta_{\lambda\sigma}
 +\delta_{\mu\sigma}\delta_{\nu\lambda}
-2\delta_{\mu\lambda}\delta_{\nu\sigma} \ne 0 $.
It may happen only when the indices $\mu,\nu,\lambda,\sigma$ may
be separated into pairs:
$(\mu_1,\mu_2)$, $(\mu_3,\mu_4)$, and $\mu_1=\mu_2$, $\mu_3=\mu_4$, $\mu_1\ne\mu_3.$
For example, $I_{\chi\chi\eta\eta}$ when $\chi \ne \eta$ diverges and needs regularization.
Let $\eta$ is equal to 1. Then with the help of the identity
\be
0=\sum_{\xi} k^{\lsb {\ssz 4}\rsb}_\xi I_{\chi\chi 1\xi}
=k^{\lsb {\ssz 4}\rsb}_0 I_{\chi\chi 10} +
 k^{\lsb {\ssz 4}\rsb}_1 I_{\chi\chi 11} +
 k^{\lsb {\ssz 4}\rsb}_2 I_{\chi\chi 12} +
 k^{\lsb {\ssz 4}\rsb}_3 I_{\chi\chi 13}
\label{WardIdCalc}
\ee
one can express $I_{\chi\chi 11}$ in terms of convergent
amplitudes $I_{\chi\chi 10}$, $I_{\chi\chi 12}$, $I_{\chi\chi 13}$
(if $k^{\lsb {\ssz 4}\rsb}_1 \ne 0$).
If $k^{\lsb {\ssz 4}\rsb}_1=0$ then another GWI's can be applied.

One sees that the set of equations~(\ref{WardIdCalc}) is sufficient
for calculating $I$ (and then $G$) with arbitrary indices.
This method simplifies calculations considerably.
The four-point function $G$ calculated with the help of this method coincides with
$G$ calculated by GPV regularization.

\section*{Comparison with the results of the dimensional regularization}

It was shown in~\cite{Dim} that there is no self-consistent
definition of $\gamma_5$-matrix in arbitrary dimension, and some
assumptions are needed for dealing with $\gamma_5$.
Let us assume that $\gamma_5=i\gamma_0\gamma_1\gamma_2\gamma_3$ in
arbitrary dimension $d$, and then
$\{\gamma_5,\gamma_\mu\}=0$ when $\mu\le 3$,
$[\gamma_5,\gamma_\mu]=0$ for others $\mu$.
The properties of $\gamma_0, \gamma_1, \ldots$ are standard:
$\{\gamma_\mu,\gamma_\nu\}=2g_{\mu\nu}$,
$\gamma_\mu\gamma^\mu=d$, ${\tr}\{1\}=d$.

Since only 'potentially divergent' part of $I$ needs regularization,
only the integrals of the following form should be examined:
\be
\int{{\rm d}\tau \int{{\rm d}^{d}p \frac
{{\tr}(\gamma_\mu {\rm p}_+ \hat p \gamma_\nu {\rm p}_+ \hat p \gamma_\lambda  \hat p \gamma_\sigma \hat p)}
{(p^2 + F(\alpha ,\beta ,\gamma ,\delta ,1,2,3,4,m^2))^4}}}
\label{dBad1}
\ee
or
\be
\int{{\rm d}\tau \int{{\rm d}^{d}p \frac
{{\tr}(\gamma_\mu {\rm p}_+ \hat p \gamma_\lambda  \hat p \gamma_\nu {\rm p}_+ \hat p \gamma_\sigma \hat p)}
{(p^2 + F(\alpha ,\beta ,\gamma ,\delta ,1,2,3,4,m^2))^4}}}
\label{dBad2}
\ee
Equations~(\ref{dBad1}) and~(\ref{dBad2}) differ by the positions
of the projectors $p_+$.
As we are interested in comparison of the dimensional
regularization with the GPV, indices $\mu, \nu, \lambda, \sigma$
are considered to have values $0,\ldots,3.$

Let us begin with~(\ref{dBad1}).
Index $\alpha$ always runs from $0$ to $3$ since there is a factor
$\ldots {\rm p}_+ \gamma_\alpha \gamma_\nu {\rm p}_+ \ldots$ in the trace.
Indeed, if $\alpha>3$ then the projector $p_+$ commutes with $\gamma_\alpha$ and since the
index $\nu \in 0,\ldots,3$ this factor is zero.
Expression~(\ref{dBad1}) splits into terms of four kinds:
1)~indices $\alpha, \beta, \gamma, \delta$ $\in 0,\ldots,3$; \,
2)~$\alpha, \beta \in 0,\ldots,3$, $\gamma, \delta \in 4,\ldots$; \,
3)~$\alpha, \gamma \in 0,\ldots,3$, $\beta, \delta \in 4,\ldots$; \,
4)~$\alpha, \delta \in 0,\ldots,3$, $\beta, \gamma \in 4,\ldots$.
These terms are denoted as $J^{1,\ldots,4}$ respectively.
For calculation of $J$'s the identity~(\ref{IntpId}) is useful.
With the help of well-known formula
$$
\int{{\rm
d}^dp\frac{(p^2)^\alpha}{(p^2+A)^\beta}}=
\pi^{d/2}A^{d+2\alpha-2\beta}\frac{\Gamma(d/2+\alpha)\Gamma(\beta-\alpha-d/2)}{\Gamma(d/2)\Gamma(\beta)}
$$
these $J$'s were calculated:
$$
J^1=\frac{d\pi^{d/2}}{6}(\delta_{\mu\nu}\delta_{\lambda\sigma}+\delta_{\mu\sigma}\delta_{\nu\lambda}-2\delta_{\mu\lambda}\delta_{\nu\sigma})
 \frac{\Gamma(2-d/2)}{F^{2-d/2}},
$$
$$
J^2=-\frac{\pi^{d/2}}{6}{\tr}(\gamma_\mu{\rm p}_+\gamma_\nu\gamma_\lambda\gamma_\sigma) \frac{\Gamma(3-d/2)}{F^{2-d/2}},
\quad
J^3=\pi^{d/2} \frac{\delta_{\mu\sigma}\delta_{\nu\lambda}}{6} \frac{\Gamma(3-d/2)}{F^{2-d/2}},
$$
$$
J^4=\frac{\pi^{d/2}}{6}(d(-\delta_{\mu\nu}\delta_{\lambda\sigma}-\delta_{\mu\sigma}\delta_{\nu\lambda}+\delta_{\mu\lambda}\delta_{\nu\sigma})
              +{\tr}(\gamma_\mu{\rm p}_+\gamma_\nu\gamma_\lambda\gamma_\sigma))
 \frac{\Gamma(3-d/2)}{F^{2-d/2}},
$$
\noindent
and $J^1+J^2+J^3+J^4=$
\be
=\left((\delta_{\mu\nu}\delta_{\lambda\sigma}+\delta_{\mu\sigma}\delta_{\nu\lambda}-2\delta_{\mu\lambda}\delta_{\nu\sigma})\Gamma(2-d/2)+
       (\delta_{\mu\lambda}\delta_{\nu\sigma}-\delta_{\mu\nu}\delta_{\lambda\sigma})\Gamma(3-d/2)
 \right)\frac{d \pi^{d/2}}{6 F^{2-d/2}}
\label{Dm1}
\ee
Let us $d=4-2\varepsilon$, then the last formula acquires the form:
$$
\frac{2\pi^2}{3}\left[(\delta_{\mu\nu}\delta_{\lambda\sigma}+\delta_{\mu\sigma}\delta_{\nu\lambda}-2\delta_{\mu\lambda}\delta_{\nu\sigma})(\frac{1}{\varepsilon} + o(\varepsilon)- \log(F) + C)
                      +(\delta_{\mu\lambda}\delta_{\nu\sigma}-\delta_{\mu\nu}\delta_{\lambda\sigma})\right],
$$
where $C$ - a constant, which depends on the chosen subtraction
scheme (for example, in the scheme of minimal subtractions $C=0$).

Calculation of~(\ref{dBad2}) is analogous.
Expression~(\ref{dBad2}) splits into terms of four kinds:
1)~$\alpha, \beta, \gamma, \delta$  $\in 0,\ldots,3$; \,
2)~$\alpha, \gamma \in 0,\ldots,3$, $\beta, \delta \in 4,\ldots$; \,
3)~$\alpha, \gamma \in 4,\ldots$,   $\beta, \delta \in 0,\ldots,3$; \,
4)~$\alpha, \beta, \gamma, \delta$  $\in 4,\ldots$
The cases 1,2,3 are considered analogously to eq.~(\ref{dBad1}).
In the fourth case integration over $p$ leads to the result:
$$
-\pi^{d/2} \frac{d-2}{12}{\tr}(\gamma_\mu{\rm p}_+\gamma_\lambda\gamma_\nu\gamma_\sigma)
 \frac{\Gamma(3-d/2)}{F^{2-d/2}},
$$
and thus the integral~(\ref{dBad2}) is equal to the following expression:
$$
\pi^{d/2} \int{\frac{{\rm d}\tau}{F^{2-d/2}}}{\scriptsize \times}
\qquad \qquad \qquad \qquad \left.\right.
$$
\be
{\scriptsize \times}
\left((\delta_{\mu\lambda}\delta_{\nu\sigma}+\delta_{\mu\sigma}\delta_{\nu\lambda}-2\delta_{\mu\nu}\delta_{\lambda\sigma})\frac{2d}{3}\Gamma(2-d/2)-
       {\tr}(\gamma_\mu{\rm p}_+\gamma_\lambda\gamma_\nu\gamma_\sigma)(1+\frac{d}{2})\frac{\Gamma(3-d/2)}{6}
 \right)
\label{Dm2}
\ee

There are six contributions into $G$ which differ by permutations.
One can note that $F(\alpha ,\beta ,\gamma ,\delta ,1,2,3,4,m^2) \le 0$
and symmetric with respect to simultaneous permutations:
$\alpha \leftrightarrow \delta,$ \,
$\beta  \leftrightarrow \gamma,$ \, and  \,
$k^{\lsb {\ssz 2}\rsb} \leftrightarrow k^{\lsb {\ssz 4}\rsb}.$
Thus there are only three different $F$'s appear in calculations.
Let us denote them as $F^{1,2,3}$:
$F^1 \equiv F(\alpha ,\beta ,\gamma ,\delta ,1,2,3,4,m^2);$\,
$\left. F^2 \equiv F^1 \right|_{k^{\lsb {\ssz 3}\rsb} \leftrightarrow k^{\lsb {\ssz 4}\rsb}};$\,
$\left. F^3 \equiv F^1 \right|_{k^{\lsb {\ssz 2}\rsb} \leftrightarrow k^{\lsb {\ssz 3}\rsb}}.$\,

Keeping in mind the symmetry properties one can write down the
contribution of all 'potentially divergent' parts into $G$:
$$
 \frac{4\pi^2}{3}\left[(\delta_{\mu\nu}\delta_{\lambda\sigma}+\delta_{\mu\sigma}\delta_{\nu\lambda}-2\delta_{\mu\lambda}\delta_{\nu\sigma})
(\frac{1}{\varepsilon} + o(\varepsilon)- \log(F^1) + C)
                      +(\delta_{\mu\lambda}\delta_{\nu\sigma}-\delta_{\mu\nu}\delta_{\lambda\sigma})\right]+
$$
$$
+\frac{4\pi^2}{3}\left[(\delta_{\mu\nu}\delta_{\lambda\sigma}+\delta_{\mu\lambda}\delta_{\nu\sigma}-2\delta_{\mu\sigma}\delta_{\nu\lambda})
(\frac{1}{\varepsilon} + o(\varepsilon)- \log(F^2) + C)
                      +(\delta_{\mu\sigma}\delta_{\nu\lambda}-\delta_{\mu\nu}\delta_{\lambda\sigma})\right]+
$$
$$
+\frac{4\pi^2}{3}\left[(\delta_{\mu\lambda}\delta_{\nu\sigma}+\delta_{\mu\sigma}\delta_{\nu\lambda}-2\delta_{\mu\nu}\delta_{\lambda\sigma})
(\frac{1}{\varepsilon} + o(\varepsilon)- \log(F^3) + C)\right. -
\qquad \qquad \qquad
$$
$$
\qquad \qquad \qquad \qquad \qquad
\left.    -\frac{3}{8}{\tr}(\gamma_\mu{\rm p}_+(\gamma_\lambda\gamma_\nu\gamma_\sigma+\gamma_\sigma\gamma_\nu\gamma_\lambda))\right]=
$$
$$
=\frac{4\pi^2}{3}\left[(\delta_{\mu\nu}\delta_{\lambda\sigma}+\delta_{\mu\sigma}\delta_{\nu\lambda}-2\delta_{\mu\lambda}\delta_{\nu\sigma})\log(F^1)
                     +(\delta_{\mu\nu}\delta_{\lambda\sigma}+\delta_{\mu\lambda}\delta_{\nu\sigma}-2\delta_{\mu\sigma}\delta_{\nu\lambda})\log(F^2)+
                     \right.
$$
$$
          \left.+(\delta_{\mu\lambda}\delta_{\nu\sigma}+\delta_{\mu\sigma}\delta_{\nu\lambda}-2\delta_{\mu\nu}\delta_{\lambda\sigma})\log(F^3)\right]+
$$
$$
+\frac{2\pi^2}{3}(\delta_{\mu\nu}\delta_{\lambda\sigma}+\delta_{\mu\sigma}\delta_{\nu\lambda}+\delta_{\mu\lambda}\delta_{\nu\sigma}).
$$
This result differs from the correct one by the constant
$\frac{2\pi^2}{3}(\delta_{\mu\nu}\delta_{\lambda\sigma}+\delta_{\mu\sigma}\delta_{\nu\lambda}+\delta_{\mu\lambda}\delta_{\nu\sigma})$.

This fact is illustrated at Fig.7, where \, $m=1;$\,
$k^{\lsb 1 \rsb}$ $=$ $(3,0,0,3);$\,
$k^{\lsb 2 \rsb}$ $=$ $(-3,0,2\sin(\theta),2\cos(\theta));$\,
$k^{\lsb 3 \rsb}$ $=$ $(3,0,0,-3);$\,
$k^{\lsb 4 \rsb}$ $=$ $(-3,0,-2\sin(\theta),-2\cos(\theta));$\,
$\theta=$ $5^o \ldots 80^o;$\,
upper line corresponds to $G_{3344}$ calculated with the help of
the dimensional regularization, middle -- to $G_{3344}$ calculated with the help of
the GPV regularization, and the lower horizontal line -- difference between them.
The accuracy of the calculations results is $\sim 10^{-7}$,
Fortran-90 and ``Maple~V'' were used.

\unitlength=1cm
\begin{picture}(14.0,7.0)
\Lengthunit=1.0cm%
\Linewidth{0.5pt}%
\put(0.5,0.8){\line(1,0){12.96}}
\put(13.46,0.8){\line(0,1){5.9}}
\put(13.46,6.7){\line(-1,0){12.96}}
\put(0.5,6.7){\line(0,-1){5.9}}
\multiput(1.22,0.8)(0.72,0){17}{\put(0,0){\line(0,1){0.15}}}
\multiput(0.5,1.3)(0,0.5){11}{\put(0,0){\line(1,0){0.15}}}
\put(0.22,1.7){\scriptsize{5}}
\put(0.10,2.7){\scriptsize{10}}
\put(0.10,3.7){\scriptsize{15}}
\put(0.10,4.7){\scriptsize{20}}
\put(0.10,5.7){\scriptsize{25}}
\put(0.42,0.53){\scriptsize{0}}
\put(6.77,0.53){\scriptsize{$\pi/2$}}
\put(13.4,0.53){\scriptsize{$\pi$}}
\put(6.0,0.02){\scriptsize{Scattering angle $\theta$.}}
\put(1.3,6.2){\scriptsize{$G_{3344}$ (dimensional regularization)}}
\put(1.3,4.0){\scriptsize{$G_{3344}$ (GPV)}}

\put(1.0,2.1159){\line(1,0){11.2}}
\multiput(0.3,0)(0,1.3159){2}%
{\put(0.0,0.0){\qbezier[20](0.7000,4.7287)(1.0520,4.6845)(1.4000,4.6351)}
\put(0.0,0.0){\qbezier[20](1.4000,4.6351)(1.7510,4.5794)(2.1000,4.5183)}
\put(0.0,0.0){\qbezier[20](2.1000,4.5183)(2.4499,4.4511)(2.8000,4.3779)}
\put(0.0,0.0){\qbezier[20](2.8000,4.3779)(3.1486,4.2993)(3.5000,4.2141)}
\put(0.0,0.0){\qbezier[20](3.5000,4.2141)(3.8473,4.1243)(4.2000,4.0274)}
\put(0.0,0.0){\qbezier[20](4.2000,4.0274)(4.5458,3.9270)(4.9000,3.8188)}
\put(0.0,0.0){\qbezier[20](4.9000,3.8188)(5.2440,3.7087)(5.6000,3.5898)}
\put(0.0,0.0){\qbezier[20](5.6000,3.5898)(5.9413,3.4713)(6.3000,3.3425)}
\put(0.0,0.0){\qbezier[20](6.3000,3.3425)(6.6377,3.2173)(7.0000,3.0793)}
\put(0.0,0.0){\qbezier[20](7.0000,3.0793)(7.3314,2.9501)(7.7000,2.8034)}
\put(0.0,0.0){\qbezier[20](7.7000,2.8034)(8.0177,2.6749)(8.4000,2.5182)}
\put(0.0,0.0){\qbezier[20](8.4000,2.5182)(8.6628,2.4096)(9.1000,2.2279)}
\put(0.0,0.0){\qbezier[20](9.1000,2.2279)(9.6184,2.0121)(9.8000,1.9368)}
\put(0.0,0.0){\qbezier[20](9.8000,1.9368)(10.196,1.7739)(10.500,1.6500)}
\put(0.0,0.0){\qbezier[20](10.500,1.6500)(10.877,1.4992)(11.200,1.3729)}
\put(0.0,0.0){\qbezier[20](11.200,1.3729)(11.569,1.2325)(11.900,1.1111)}}
\end{picture}
\begin{center}
\qquad Fig.7.
\end{center}

This difference is due to the presence of exactly two projectors
in the traces in equations~(\ref{dBad1}),(\ref{dBad2}),
because the gluons are transformed by the vectorial $\rm SU(3)$ group,
whereas $W$--mesons are transformed the chiral $\rm SU(2)$ groups.
The correct result can be recovered with the help of GWI,
introducing the finite counterterm.

Hence a straightforward application of the dimensional
regularization with $\gamma_5$ defined as above breaks gauge
invariance of the model.

\section*{Discussion}

The comparison of different regularizations methods for the four-point function is carried
out. The contribution of the spinor loop to the four-point function for $W^+W^- \to gg$ transition
in the Standard model is
examined by means of the Generalized Pauli-Villars
regularization, dimensional regularization and with the help of
Generalized Ward Identities.

It is found that GPV regularization and renormalization with the
help of the GWI give correct result, whereas
inconsistency in the definition of $\gamma_5$-matrix in the framework of the
dimensional regularization leads to the result, which differs from the correct one
by a constant.

For the diagram we considered, all three methods require a
comparable amount of work. However for more complicated diagrams
the manifestly gauge invariant GPV regularization seems to be
preferable.

\section*{Acknowledgements}

One of the authors (A.A.S.) thanks D.Dyakonov for the useful
discussion which stimulated this study.

This work was supported in part by RBRF under grant 99-01-00190
and by the grant 96208 for support of leading scientific schools.

$$~$$

\end{document}